\documentclass[12pt]{article}
\usepackage{epsfig}
\usepackage{graphicx}
\usepackage{cite}
\usepackage{amsfonts}
\usepackage{amssymb}
\usepackage{latexsym}
\def\be{\begin{equation}}
\def\ee{\end{equation}}
\def\ba{\begin{eqnarray}}
\def\ea{\end{eqnarray}}

\def\bq{\begin{quote}}
\def\eq{\end{quote}}

 at 10truept

\newcommand{\beq}{\begin{equation}}
\newcommand{\eeq}{\end{equation}}
\newcommand{\beqa}{\begin{eqnarray}}
\newcommand{\eeqa}{\end{eqnarray}}
\newcommand{\bea}{\begin{eqnarray}}
\newcommand{\eea}{\end{eqnarray}}
\newcommand{\p}{\partial}

\def\lesssim{~\mbox{\raisebox{-.6ex}{$\stackrel{<}{\sim}$}}~}

\def\ltap{\ \raise.3ex\hbox{$<$\kern-.75em\lower1ex\hbox{$\sim$}}\ }
\def\gtap{\ \raise.3ex\hbox{$>$\kern-.75em\lower1ex\hbox{$\sim$}}\ }
\def\gl{\ \raise.5ex\hbox{$>$}\kern-.8em\lower.5ex\hbox{$<$}\ }
\def\roughly#1{\raise.3ex\hbox{$#1$\kern-.75em\lower1ex\hbox{$\sim$}}}

\def\half{\frac{1}{2}}
\def\ZZ{\mathbb{Z}}

\renewcommand{\thefootnote}{\fnsymbol{footnote}}

\begin{document}

\thispagestyle{empty}
\begin{flushright}
BRX-TH-6306\\
July 2016
\end{flushright}
\vspace*{.05cm}
\begin{center}
{\Large \bf A Monodromy from London}

\vspace*{.75cm} {\large Nemanja Kaloper$^{a, }$\footnote{\tt
kaloper@physics.ucdavis.edu} and  Albion Lawrence$^{b, }$\footnote{\tt
albion@brandeis.edu}}\\
\vspace{.7cm} {\em $^a$Department of Physics, University of
California, Davis, CA 95616, USA}\\
\vspace{.3cm} {\em $^b$Martin Fisher School of Physics, Brandeis University, Waltham, MA 02453, USA}\\

\vspace{1.5cm} {\bf Abstract}
\end{center}
We focus on the massive gauge theory formulation of axion monodromy inflation. We argue that 
a gauge symmetry hidden in these models is the key mechanism protecting inflation
from dangerous field theory and quantum gravity corrections. The effective theory of large field inflation is dual to a massive U(1) 4-form gauge theory, which is similar to a massive gauge theory description of superconductivity. The gauge theory explicitly realizes the old Julia-Toulouse proposal for a low energy description of a gauge theory in a defect condensate. While we work mostly with the example of quadratic axion potential induced by flux monodromy, we discuss how other types of potentials can arise from inclusion of gauge invariant corrections to the theory.

\vfill \setcounter{page}{0} \setcounter{footnote}{0}
\newpage

\renewcommand{\thefootnote}{\arabic{footnote}}
\setcounter{equation}{0} \setcounter{footnote}{0}

\setcounter{tocdepth}{2}
\tableofcontents

\section{Introduction}

Large field models of inflation, in which the canonically normalized inflaton $\phi$ ranges over a distance $\geq m_{pl}$ in field space, have an appealing simplicity \cite{Linde:1983gd,Freese:1990rb}: they can be built from single field with a flat and even monomial potential, and some direct couplings to matter that the inflaton decays into at the end of inflation, reheating the universe. They generate the largest possible primordial tensor fluctuations, which could be detected by the new era of CMBR polarization experiments sensitive to gravitational radiation.

The challenge such models present is that the infinite number of terms of the form 
\be 
	\delta L \sim c_n {\phi^n}/{m_{pl}^{n-4}}\label{eq:annoying}
\ee
must have exquisitely small coefficients $c_n$ lest slow roll and vacuum energy dominance be spoiled. This is not a problem in perturbative quantum field theory, where such models can be radiatively stable, so long as $\phi$ is derivatively coupled to any heavy physics that it may reheat into. Even when graviton loops are included, models with small initial values of $c_n$ are still technically natural \cite{Linde:1987yb}, because an approximate shift symmetry \cite{Kaloper:2008fb,Kaloper:2011jz}\ becomes exact when $c_{n>0} = 0$.  On the other hand, there is a body of theoretical evidence that quantum gravity does not allow for unbroken or weakly broken global symmetries, bringing into question the initial choice $c_n \ll 1$. Some mechanism which restricts the UV theory is required.

Axion monodromy inflation \cite{Silverstein:2008sg,McAllister:2008hb,Kaloper:2008fb,Kaloper:2011jz}\ provides a candidate mechanism for UV completion of large field inflation. Most work on these models is by necessity rather technical, as it involves explicit string theory constructions in order to discuss the UV-complete theory in a controlled fashion. Our goal here is to understand the field theory description of what the successful constructions lead to. We will argue that monodromy inflation models use a hidden gauge symmetry which ensures the consistency of the low energy theory and protects it from large corrections descending from the ultraviolet (UV) completion.

More specifically, the low-energy theory can be written as a massive $U(1)$ 4-form gauge theory \cite{Kaloper:2008fb,Kaloper:2011jz,Kaloper:2008qs,Marchesano:2014mla}, with the axionic inflaton dual to the longitudinal mode of the 4-form.   Such massive gauge theories can arise from the old Julia-Toulouse description of phases which contains many on-shell topological defects \cite{Julia:1979ur,Quevedo:1996uu,Quevedo:1996tx}. We will review it and provide a new argument for the core ideas in \cite{Julia:1979ur,Quevedo:1996uu,Quevedo:1996tx}. The phases 
with condensed defects 
are not continuously connected to the usual perturbative vacuum of the original theory. The large inflaton field excursions encode {\it macroscopic} field configurations: large 4-form flux fluxes induced by a condensate of topological defects. We will show that this theory is well behaved at high energies, so that the UV completion enters only through irrelevant operators. The macroscopic nature of the field configurations ensures that these irrelevant operators do not spoil slow-roll inflation. The upshot is a London equation-level description of monodromy inflation, which complements the BCS-type constructions in \cite{Silverstein:2008sg,McAllister:2008hb,Kaloper:2008fb,Kaloper:2008qs,Marchesano:2014mla,DAmico:2012ji,DAmico:2012sz,DAmico:2013iaa}. The underlying effective theory is analogous to the massive gauge theory description of the ground state of a superconductor. This picture also sheds some light on the fate of the Weak Gravity Conjecture \cite{ArkaniHamed:2006dz}\ for massive gauge fields.

We avoid the problems stemming from the sad fate of global shift symmetries in quantum gravity because there is no global shift symmetry to break. The discrete shift symmetry of the dual axion $\phi$ is not a remnant of a broken global symmetry, but a gauge symmetry which emerges from the dual gauge symmetry of the massive 4-form. As shown in \cite{Kaloper:2011jz}, this prohibits direct contributions of the form (\ref{eq:annoying}). One still needs to ensure that the construction of the massive U(1) 4-form gauge theory is consistent with the UV completion, but that is a more straightforward task, which should be addressed with the existing tools.

While we focus here on generating quadratic axion potentials, we also discuss extensions which display flatter potentials, which emerge in the original string theory constructions \cite{Silverstein:2008sg,McAllister:2008hb,Dong:2010in,McAllister:2014mpa}. These can arise from coupling to additional fields controlled by sub-Planckian dynamical scales.

With this motivation in mind, we will first discuss massive $U(1)$ vector fields, to highlight the essential physics in a more familiar context. In particular we will give a novel, and we feel compelling, motivation for the Julia-Toulouse description of the hydrodynamics of vortices.

\section{Warmup: massive $U(1)$ vector fields}

\subsection{Background}

Massive U(1) gauge theory contains three propagating degrees of freedom: the
standard two transverse modes, and the additional longitudinal mode. These obey a massive dispersion relation $p^2 + m^2 = 0$ on shell. The theory has a nonlinearly realized gauge symmetry. We will describe this directly below, but an intuitive way to see this is to consider the massive theory as the low-energy limit of a weakly coupled Abelian Higgs model in the symmetry breaking phase.  The mass term arises from mixing between the photon and the Goldstone mode of the charged condensate. This Goldstone mode acts as the longitudinal mode of the photon, transforming nonlinearly under the $U(1)$ gauge symmetry. 

A well-known manifestation occurs in superconductors governed by BCS theory.  The condensate of Cooper pairs at very low energies is decoupled from other excitations of the medium due to the mass gap. The long distance correlation between the Cooper pairs implies that the quantum-mechanical phase invariance is spontaneously broken in the material by the condensate. Hence there is a dissipationless relativistic sound wave, by the Goldstone theorem, which sustains a background current because the Cooper pairs are charged. The current persists even in the stationary limit, and the vector potential does not vanish but is proportional to it,
\be
\vec A = \frac{ \vec J}{m^2} \, .
\ee
This is London's equation, which is the basis for phenomenological theory of superconductivity. Substituting this expression for $J$ into Maxwell's equations yields the equations of motion for a massive photon. 

The massive equation of motion can be easily obtained as the non-relativistic limit of the Poincar\'e invariant Proca Lagrangian\footnote{Note that we use the "east coast" signature convention $(-+++)$ in this paper.}
\be \label{proca}
{\cal L} = - \frac14 F_{\mu\nu}^2 - \frac12 m^2 A_\mu^2 + A_\mu J^\mu \, .
\ee
Here $F_{\mu\nu} = \partial_\mu A_\nu - \partial_\nu A_\mu$, and $J^\mu$ is the conserved current of the charged matter, $\partial_\mu J^\mu=0$. Varying the equation with respect to $A$ yields
\be
	\Box A_{\mu} - \p_{\mu} \p \cdot A - m^2 A_{\mu} = J_\mu\ .
\ee
Taking the divergence of this equation yields $\p\cdot A = 0$ for $m^2 \neq 0$: the Lagrangian (\ref{proca}) is already gauge-fixed.

Current conservation is required for internal consistency of the theory, to guarantee the absence of ghosts.\footnote{One can introduce current non-conservation by taking some very heavy charged matter and integrating it out. The processes which involve both light and heavy charged matter would lead to unitarity violations and current nonconservation at low energies, suppressed by the mass scale of the heavy matter. The low energy theory would appear to have a problem with ghosts. However the ghost mass would be comparable to the mass of the heavy matter.  Unitarity is restored at that scale by integrating the heavy matter back in.} It also guarantees that (\ref{proca}) is perfectly well behaved at the tree level even in the $m \to 0$ limit.  

The gauge implicit in (\ref{proca})\ is not ideal for examining the high-energy behavior of the theory.  The massive photon propagator derived from (\ref{proca}) is
\be \label{procaprop}
\Delta_{\mu\nu} = \frac{i}{p^2 + m^2} \left( \eta_{\mu\nu} + \frac{p_\mu p_\nu}{m^2} \right) \, ,
\ee
and features singular behavior as $m^2 \rightarrow 0$, due to the second term in the parenthesis.  Na\"ively this term could appear in processes with virtual photons. In fact, as long as the external lines involve only conserved currents, $p_\mu J^\mu = 0$, the contributions from this term to any physical S-matrix element vanish identically, suggesting that the singularity is merely a gauge artifact, stemming from the gauge choice $\partial \cdot A = 0$. The next question is whether this theory has a good $m^2 \to 0$ limit when one considers loops containing virtual photons, since those involve contractions of the momentum factors in terms $\propto p_\mu p_\nu/m^2$ off shell, which do not vanish trivially. Here, gauge redundancies come to the rescue, as we will now review. 

\subsection{Renormalizability}

Massive Abelian gauge theories coupled to conserved currents are known to be consistent renormalizable quantum theories with a smooth massless limit.  This is reviewed in detail in many textbooks.  We will give a heuristic discussion of the underlying physics which keeps the theory healthy.

Inspired by BCS theory, we change the gauge implied by (\ref{proca}) and separate the longitudinal scalar mode from $A_{\mu}$. To this end we shift $A_\mu \rightarrow A_\mu - \p_\mu \phi$, and treat $\phi$ as a field transforming nonlinearly under the gauge symmetry, as per Stueckelberg's trick (note that $\phi$ is dimensionless). The Lagrangian (\ref{proca}) becomes manifestly gauge invariant.  Finally, we add a gauge fixing term, in order to properly quantize the theory, and arrive at the Lagrangian:
\be \label{stuckelberg}
{\cal L} = - \frac14 F_{\mu\nu}^2 - \frac12 m^2 (A_\mu - \p_\mu \phi)^2 + ( A_\mu - \p_\mu \phi) J^\mu - \frac{1}{2\alpha} (\p \cdot A - \alpha m^2 \phi)^2 \, .
\ee
The parameter $\alpha$ is an arbitrary gauge fixing parameter. The scalar-current coupling $\p_\mu \phi \, J^\mu$ is just a boundary term when the current is conserved and can be ignored henceforth: we only keep it to make the first three terms manifestly gauge invariant under $A_\mu \rightarrow A_\mu + \p_\mu \Omega$, $\phi \rightarrow \phi + \Omega$. The gauge fixing term imposes the $R_{\xi}$-like gauge 
\be 
	\p \cdot A - \alpha m^2 \phi = 0\ .\label{eq:rxi}
\ee
Our gauge choice removes the bilinear cross term $\propto m^2 A^\mu  \, \p_\mu \phi$, ensuring that the vacuum is manifestly stable, and simplifying the propagators. 
 
The gauge (\ref{eq:rxi})\  follows by applying a gauge transformation $\Omega$ to arbitrary $A'_\mu$ and $\phi'$, such that $\Omega$ satisfies $(\p^2 - \alpha m^2) \, \Omega = - (\p \cdot A' - \alpha m^2 \phi')$. The transformed fields satisfy the gauge condition (\ref{eq:rxi}). The theory retains a residual gauge symmetry 
\be
	A_\mu \rightarrow A_\mu + \p_\mu \omega\ ,\ \ \ \ \ \phi \rightarrow \phi + \omega\ ,\ \ \ \ \ (\p^2 - \alpha m^2) \, \omega = 0\ .\label{eq:resid}
\ee
The transformation (\ref{eq:resid})\ preserves both the field equations and the gauge choice. In covariantly quantized perturbation theory, the gauge condition is imposed on the Hilbert state by requiring that they are in the kernel of the positive frequency part of the gauge fixing condition, $(\p \cdot A - \alpha m^2 \phi)^{(+)} |phys \rangle = 0$.

While we appear to have five degrees of freedom -- the four components of the gauge field, and the scalar -- two scalar modes are redundant. The fields $\p \cdot A$ and $\phi$ satisfy the same field equation thanks to current conservation. They and the residual gauge mode $\omega$ reside on exactly the same mass shell, which for general $\alpha$ is not degenerate with the vector modes. The residual transformation (\ref{eq:resid})\   can be used to set either of the two scalars $\p \cdot A, \phi$ to zero. The gauge-fixing condition $\p \cdot A - \alpha m^2 \phi = 0$ then sets the other field to zero.  Three degrees of freedom remain in the theory (\ref{stuckelberg}), living on the physical mass shell $p^2 = - m^2$.\footnote{This can be seen by projecting the mode $\p\cdot A$ out of the gauge field $A^{\mu}$: see for example \S3-2-2\ in \cite{Itzykson:1980rh}.} The theories (\ref{proca}) and (\ref{stuckelberg}) are thus exactly the same.

The theory (\ref{stuckelberg}) has manifestly healthy behavior on and off shell in the limit $m^2 \rightarrow 0$, which is equivalent to the limit $p^2 \rightarrow \infty$ at fixed $m$. To see this, we write the momentum space propagators following from (\ref{stuckelberg}) after canonically normalizing the scalar by $\phi = \varphi/m$:
\be\label{stuckprops}
\Delta_{\mu\nu} =  \frac{i}{p^2 + m^2} \left( \eta_{\mu\nu} + (1-\alpha) \frac{p_\mu p_\nu}{p^2 + \alpha{m^2}}  \right) \, ,  ~~~~~~~~~~~
\Delta_{\varphi} =  - \frac{i}{p^2 + \alpha m^2} \, .\ee
Consistent with the discussion above, the structure of these propagators shows that the theory has in effect two mass shells, one at $p^2 + m^2 = 0$ where the three physical massive photon polarizations reside, and another at $p^2 + \alpha m^2 = 0$, where the two extra scalars dwell. The latter modes are unphysical since their mass depends explicitly on the gauge fixing parameter $\alpha$. Indeed, as noted above, they can be removed by the residual gauge transformation $\omega$. Their sole role is to allow for breaking the degeneracy between the physical and unphysical sectors, to show that the theory is manifestly well behaved at high momenta. For finite $m^2, \alpha m^2$, the propagators now vanish as $p^2 \rightarrow \infty$, implying that the theory remains weakly coupled all the way to UV, and that all the S-matrix elements at an arbitrary loop level remain finite and well behaved.  

The problematic terms in the Proca gauge propagator (\ref{procaprop}) come from the contribution $\propto p_\mu\ p_\nu/m^2$ in the parenthesis. In (\ref{stuckprops})\ they are replaced by
$\propto p_\mu p_\nu/(p^2 + \alpha m^2)$, which saturate in the UV to $\simeq p_\mu p_\nu/p^2$, so that the propagator reduces to the massless propagator at leading order in $m^2/p^2$. The contributions from  the photon mass in the loop diagrams are just IR corrections. There are no terms $\propto p^2/m^2$ in the UV that could spoil the UV behavior of the theory, because the Stueckelberg scalars are free fields.  The singular contribution $\propto p_\mu p_\nu/m^2$ in the gauge (\ref{proca}) is pure gauge, induced by the fact that the scalar field's charge is $\propto m$, and that $\phi$ is not canonically normalized.  The theory is therefore renormalizable for any $\alpha$ by simple power counting, as long as the current $J^\mu$ is conserved. 

Furthermore, there are no divergent counterterms proportional to mass term $ m^2 A_\mu^2/2$ \cite{ruegg}. All of the counterterms that would be required for renormalizing the vector sector of the theory come from the divergences in the photon self-energy. Since the photon, massive or not, is neutral, and the Stueckelberg scalars are free fields, the photon self-energy must involve at least one matter current operator insertion. As we demand current conservation, the photon self-energy in momentum space obeys $p_\mu \Pi^{\mu\nu} = 0$. Together with Poincar\'e symmetry, this implies $\Pi_{\mu\nu} = (\eta_{\mu\nu} p^2 - p_\mu p_\nu) \, \Pi(p^2)$.  
$\Pi(p^2)$ must be dimensionless.  If the matter sector is renormalizable, then by power counting $\Pi$ has at most a logarithmic divergence; if the matter sector contains irrelevant operators suppressed by a high cutoff $\Lambda$, $\Pi$ will still grow logarithmically up to this scale $\Lambda$.  Since $\Pi_{\mu\nu}$ is proportional to the external momentum, the only counterterm required is a wavefunction renormalization term.  This will induce a logarithmic running of the photon mass term (which in general could arise from both mass and wavefunction renormalization counterterms), but the shift will be proportional to $m^2$. Thus the mass term is radiatively stable.

We close with a scrutiny of the $m^2 \rightarrow 0$ limit of the pure massive gauge theory. For fixed $\alpha$, the propagators (\ref{stuckprops}) become
\be\label{stuckpropslim1}
\Delta_{\mu\nu} = \frac{i}{p^2} \left( \eta_{\mu\nu} + (1-\alpha) \frac{p_\mu p_\nu }{p^2}  \right) \, ,  ~~~~~~~~~~~
\Delta_{\varphi} =  - \frac{i}{p^2} \, , \ee
which are just the propagators of massless QED with a gauge fixing term $\frac{1}{2\alpha} (\p \cdot A)^2$, together with a canonically normalized free scalar field with Lagrangian $\frac12 (\p \varphi)^2$. Nothing is charged under the scalar, due to gauge symmetry and renormalizability of the original massive $U(1)$ theory. There are no long range forces that it can mediate. So the scalar is completely fixed by the classically imposed boundary conditions, and can never change in any quantum process. In flat space, it can simply be set to zero, since it is completely unphysical.  

There is an apparent subtlety when gravity is turned on, since in this limit the field $\varphi$ has a nonvanishing stress-energy tensor. 
While this field can be gauged away from the stress energy tensor for any nonvanishing $m^2$, the canonically normalized field transforms as $\varphi \to \varphi + m \Omega$, and so becomes gauge invariant in the massless limit, coupling to gravity as an additional minimally coupled scalar. The zero mode of the scalar is frozen out (its periodicity scales as $m/e$); momentum modes couple to gravity, but to nothing else. The only effect will arise from this scalar running in loops and correcting the gravitational effective action. This is an unobservable, scheme-dependent effect: at worst it will be absorbed into the definition of the bare couplings.

One may have worried about this in the context of the Weak Gravity Conjecture (WGC) \cite{ArkaniHamed:2006dz}, since the scalar can be written as a dual 2-form gauge field with vanishing gauge coupling.  So long as we disallow any string solutions with conserved charge under this scalar, the argument of \cite{ArkaniHamed:2006dz}\ does not apply.  The presence of such strings indicates a UV completion in which the longitudinal mode couples to additional degrees of freedom and the mass is no longer elementary, as happens in the Abelian Higgs model. In that case the massless limit above yields a gauge theory together with a massless charged scalar. Such a theory automatically satisfies the lower bound on charged masses given in \cite{ArkaniHamed:2006dz}.

\subsection{The low-energy effective action for massive gauge fields}

In any scenario we care about, our massive gauge theory will be embedded in a more elaborate field theory or string theory with both additional light fields and new dynamics above some heavy mass scale $\Lambda$.  The avatars of this high-scale dynamics will be the renormalization of operators in the low-energy theory.  Our goal is to perform an effective field theory analysis to deduce what operators can arise, and what their strength can be. Since the low-energy theory is renormalizable, with a nonlinearly realized gauge symmetry, the effective field theory is constrained. 

The mass term receives no divergent counterterms, and the gauge charge will have at most logarithmic divergences.  The remaining na\"ively renormalizable operator is $A^4$, which we will discuss below as an example within a family of operators. The remaining operators are irrelevant, and since  massive gauge theory is renormalizable, these operators will be generated only by integrating out the UV fields, suppressed at least by the powers of $\Lambda$ determined by dimensional analysis.

The simplest case are the operators of the form $F^{2k}$. These are not prohibited by any symmetry, and will take the form ${F^{2k}}/{\Lambda^{4k-4}}$ with ${\cal O}(1)$ dimensionless couplings.  

The more interesting set of terms are those with the form
\be \delta {\cal L} \sim c_k A^{2k} \, . \label{eq:Apowers}
\ee 
These carry the same nonlinearly realized gauge symmetry that the mass term in (\ref{proca}) does, so they are not  forbidden. Nonetheless, we claim that they are suppressed compared to what simple dimensional analysis would suggest: if
\be
	c_k = \frac{{\tilde c}_k}{\Lambda^{2k-4}} \, , \label{eq:dlcoeff}
\ee
then
\be
	{\tilde c}_k \sim d_k \left(\frac{m}{\Lambda}\right)^{2k} \, , \label{eq:mcoeff}
\ee
where $d_k$ is of order ${\cal O}(1)$ or at worst ${\cal O}\left(\ln (\Lambda/m)\right)$. 
For $k = 2,4$ the operators are relevant and marginal by na\"ive power counting.  In the case $k = 2$, we have already shown that this is true for the mass term itself.  We might expect that a counterterm could be generated for $k = 4$; however, the arguments below for general $k > 2$ will show that this will be suppressed by a factor of $m^4$.

In a general $R_{\xi}$ gauge the gauge field propagator at high energies approaches that of the massless propagator. Thus, the high energy behavior which controls counterterms will be dominated by the massless theory. This is a consequence of gauge symmetry. We thus expect that terms of the form (\ref{eq:Apowers})\ should vanish as $m^2 \to 0$ and the manifest gauge symmetry is restored. The question is the power of $m$ that should govern $c_k$. 

At this point we will assume that the UV completion of our theory is either a renormalizable QFT, or a theory such as string theory with soft high-energy behavior. These theories have normal decoupling behavior at high energies, such that when $p^2 \gg m^2$, physical processes receive only small corrections from the gauge field mass. Furthermore, in the energy range $m \ll E \ll \Lambda$, S-matrix elements will grow at most as $\ln(E)$. We will also keep the Stueckelberg mode $\phi$ explicit.  This field could descend from the angular mode of a charged scalar in the UV completion, or as a phase of some fermion condensate.

With this assumption, we can appeal to the Goldstone Boson Equivalence Theorem (GBET) \cite{Lee:1977eg,Cornwall:1974km,Chanowitz:1985hj,Kilgore:1992re}, which governs the behavior of irrelevant operators in effective field theories with massive gauge particles, when such theories arise as low-energy limits of renormalizable QFTs.  
We will base our discussion in particular on the discussion and proof in \cite{Chanowitz:1985hj}. Using gauge invariance in the form of BRST invariance in $R_{\xi}$ gauge, the theorem states that the S-matrix for $n$ longitudinal gauge bosons is equivalent to the S-matrix for $n$ scalars $\phi$, up to terms which are powers of $m/E$ (perhaps times $\ln E$ terms). At such high energies, the longitudinal polarization vector has the form 
\be
	e_{\mu,L} \sim \frac{k_\mu}{m} + {\cal O}(\frac{m}{E})\ .
\ee
If we write an effective action whose coefficients generate these S-matrix elements from tree-level diagrams, terms of the form $m^{2k} A^{2k}$ will have the same coefficients as terms of the form $(\p\varphi)^{2k}$. Note that the GBET guarantees this even if we introduce direct couplings of the Stueckelberg mode to other degrees of freedom via non-current interactions or irrelevant operators, as long as these couplings do not explicitly break gauge symmetry.

Finally, we demand that the S-matrix elements for $\phi$ should not diverge as $m^2 \to 0$, beyond the usual IR divergences which can be dealt with by properly specifying the long wavelength observables. Thus, the coefficients of $(\p\varphi)^{2k}$ in the low energy effective action should be constant or scale with positive powers of $m$. If the gauge field mass is not elementary, this simply means that the Goldstone mode must arise from a physical field which conforms with our assumptions above. 
This demand combined with the GBET then implies our claim about mass dependence of the terms of the form $A^{2k}$. Gauge symmetry and absence of singularities in the effective action for $A$ in the massless limit are sufficient to show that the coefficients of the irrelevant operators $\propto A^{2k}$ can only involve the gauge field mass in numerators. The argument above fixes the minimum positive power of $m$ to be that given in (\ref{eq:Apowers}--\ref{eq:mcoeff}).  

The proof in \cite{Chanowitz:1985hj}\ made the more restrictive assumption that the UV theory was a renormalizable field theory.  We wish to reassure the reader that our less restrictive assumptions  should not spoil the theorem. While string theory is not of course a renormalizable 4d field theory in the usual sense, the irrelevant operators it generates will be suppressed by a high scale such as the string or Planck scale.  So long as this scale is at or above the scale $\Lambda$, the argument above goes through. The additional operators with dimension $\Delta$ will contribute terms scaling as $(E/\Lambda)^{\Delta-4}$ to S-matrix elements, which remain small in the limits we are considering when $\Delta > 4$. Of course, this is also the reason that one can ignore Planck-suppressed operators when using the GBET to study electroweak dynamics.

\subsection{Julia-Toulouse and generation of mass}

The `conventional' UV completion of a massive gauge theory is via the Higgs mechanism, in which an electrically charged field condenses, spontaneously breaking the charge symmetry. The Goldstone mode of the broken symmetry then becomes `eaten' by the gauge field because the symmetry is gauged, and becomes the longitudinal mode of the gauge field.

Another route was suggested by Julia and Toulouse in the late '70's \cite{Julia:1979ur}, who proposed that the correct hydrodynamic variable for a fluid of vortices was a massive gauge field.  This argument was then generalized by Quevedo and Truegenberger \cite{Quevedo:1996uu,Quevedo:1996tx} to higher rank gauge theories, who also discussed some of the aspects of the dynamical origin of the Julia-Toulouse proposal. Some studies of specific microscopic models of condensation were pursued recently in \cite{Grigorio:2011pi,Grigorio:2012jt}.
In this section we will review this proposal and provide a novel argument for it, which complements the prior work and, we hope, sheds some light on this fascinating proposal.

Consider a compact scalar field $\phi$ in a state which contains string-like topological defects.  The string dynamics are governed by some UV completion which we will not specify, other than that this dynamics supports a state in which there are a large number, or a condensate of such strings, justifying a coarse-grained hydrodynamic description.  

Outside of the vortex cores, $d\phi$ is a closed but not exact form, as $\phi$ is only defined up to shifts by its periodicity. Inside the vortex core, in the UV-complete theory, $d\phi$ will not even be a closed form due to the interactions that support the string configuration. 

Julia and Toulouse argue for their proposal by noting that the phase $\phi$ varies in a system of adiabatically moving vortices according to $\delta \phi \sim \vec v \cdot \delta \vec x$. Since in presence of vortices the fluid is not irrotational, the velocity is not a gradient of the phase, $\vec v \ne \vec \nabla \phi$, but is instead leading to an independent superfluid current vector $\vec j \propto \vec v$. They then promote the  superfluid current $\vec j$ to a gauge field, $\vec V \sim \vec j$. This is London's equation, which implies that the underlying effective gauge theory must be massive. We provide an alternative argument here. 

Outside of the vortex cores, we can perform Abelian duality, mapping the compact scalar $\phi$ to a two-form gauge potential $B_{\mu\nu}$ with 3-form field strength $H = dB$.  This is just the standard Kalb-Ramond 3-form field strength. The vortices will couple electrically to $B$. It is natural to suppose that when the vortices condense and yield a background current which mixes with $B$, the low-energy theory is described by a massive 2-form potential. This will be our key assumption. We then separate out the longitudinal modes of $B$ (which model the electrically charged vortex current outside of the vortices) and treat them as Stueckelberg fields, with the Lagrangian 
\be \label{jt3}
	{\cal L}_{H} = - \frac{1}{12} H_{\mu\nu\lambda}^2 - \frac{m^2}{4} (B_{\mu\nu} - \frac{1}{m} {{\tilde F}_{\mu\nu}})^2 + \ldots \, .
\ee
Here ${\tilde F}_{\mu\nu}$ is the field strength for a one-form Stueckelberg potential ${\tilde A}_{\mu}$, and we have chosen our normalization for later convenience.  The nonlinearly realized gauge symmetry for $B$ is
\be
	B \to B + d\Lambda^{(1)}\, ,  ~~~~~~~~~~~~  {\tilde A} \to {\tilde A} + m \Lambda^{(1)} \, ,
\ee
where $\Lambda^{(1)}$ is a one-form gauge transformation. Note that we have written the Lagrangian (\ref{jt3}) in Proca-like gauge, in which the duality transformations we describe below are most straightforward.

The low-energy dynamics of the vortex condensate is parametrized by the Stueckelberg ``current" ${\tilde F}$, which mixes with $B$. The mass $m$ arises as an IR parameter of the condensed phase. It is determined by the detailed microphysics that controls the core of the defects and thus the defect-defect interactions, just as the gauge field mass in a superconductor depends on the detailed microphysics driving the formation and condensation of Cooper pairs. Once it is established that the defects do condense, leading to (\ref{jt3}) as the low-energy theory, the mass $m$ receives no divergent contributions, so that the massive phase generated by the defect condensation is not destabilized by small changes in the dynamics of the microscopic theory.

To arrive at the Julia-Toulouse description, let us now dualize in the other direction.  If we start from (\ref{jt3}), $B$ is hard to dualize directly, as duality works at the level of field strengths. But we can start by applying the duality to $\tilde A$, so long as we work in a Proca-like gauge in which ${\tilde A}$ appears only via its field strength.  To this end consider the Lagrangian
\be
	{\cal L}_{BF} = - \frac{1}{4} F_{\mu\nu}^2 -  \frac{1}{12} H_{\mu\nu\lambda}^2 - \frac{m}{4} \epsilon_{\mu\nu\lambda\sigma} B^{\mu\nu} F^{\lambda\sigma} + \frac{1}{2}  \epsilon^{\mu\nu\lambda\sigma} {\tilde A}_{\mu} \, \p_\nu F_{\lambda\sigma} + \ldots \, ,
\ee
where we introduce a new gauge field strength $F_{\mu\nu}$. If we integrate the last term by parts, and integrate the Gaussian field $F$ out of the path integral, we arrive precisely at (\ref{jt3}). On the other hand, $\tilde A$ now appears manifestly as a Lagrange multiplier. Hence we can integrate it out of the path integral,  enforcing the Bianchi identity for $F$, which implies that there exists a new gauge field potential $A_\mu$ such that $F_{\mu\nu} = \p_\mu A_\nu - \p_\nu A_\mu$. This leaves us with the $B\wedge F$ action of a topologically massive gauge theory:
\be
	{\cal L}_{BF} = - \frac{1}{4} F_{\mu\nu}^2 -  \frac{1}{12} H_{\mu\nu\lambda}^2 - \frac{m}{4} \epsilon_{\mu\nu\lambda\sigma} B^{\mu\nu} F^{\lambda\sigma} + \ldots \, . \label{eq:bf}
\ee
Finally, we can perform one more duality transformation: we take $H$ and map it back to the original scalar $\phi$. Proceeding similarly as above, we define a new gauge field strength and potential,  $V_{\mu\nu} = \p_\mu V_\nu - \p_\nu V_\mu $, and consider the action
\be \label{stuckjt1}
	{\cal L}_{JT} =  -\frac{1}{4} F_{\mu\nu}^2 - \frac{m^2}{2}  (A_\mu - V_\mu)^2 - \frac{m}{4} \epsilon^{\mu\nu\lambda\sigma} B_{\mu\nu} V_{\lambda\sigma} + \ldots \, .
\ee
If we integrate the final term by parts and integrate the field $V$ out of the action, we arrive at (\ref{eq:bf}). On the other hand, $B_{\mu\nu}$ now appears as a Lagrange multiplier. If we integrate over it in the path integral we enforce $V_{\mu\nu} = 0$, which implies that locally the 
gauge field $V_\mu$ is pure gauge, $V_{\mu} = \p_{\mu} \phi$. Substituting this into the action, we finally arrive at the massive vector gauge theory action in Stueckelberg form,
\be \label{stuckjt}
	{\cal L}_{JT} =  -\frac{1}{4} F_{\mu\nu}^2 - \frac{m^2}{2}  (A_\mu - \p_\mu \phi)^2 + \ldots \, .
\ee
So as we said above, we see that the scalar $\phi$ is the Stueckelberg field for a massive gauge field $A_{\mu}$, which is precisely the field Julia and Toulouse proposed to be the variable to describe the hydrodynamics of superfluid vortices. The precise duality relationship between these fields is
$A_\mu - \p_\mu \phi = \frac{1}{6m} \epsilon_{\mu\nu\lambda\sigma} H^{\nu\lambda\sigma}$, which after fixing to the Proca gauge $A'_\mu = A_\mu - \p_\mu \phi$ becomes
\be\label{jtansatz}
A'_\mu = \frac{1}{6m} \epsilon_{\mu\nu\lambda\sigma} H^{\nu\lambda\sigma} \, .
\ee
We can see through this chain dualities that $A$ is dual to the gauge field ${\tilde A}$. The latter gives the Goldstone-like dynamics of a condensate of charged membranes. The Julia-Toulouse field $A_{\mu}$ is dual to that Goldstone mode, while the longitudinal component is dual to the gauge field $B$ which eats the Goldstone mode. We feel that this demystifies the Julia-Toulouse ans\"atz.

Quevedo and Truegenberger \cite{Quevedo:1996uu,Quevedo:1996tx}\ also argued that the Julia-Toulouse mechanism was dual to the Higgs mechanism. Their arguments began with writing actions for $p$-form gauge fields coupled to both electric and magnetically charged defects with delta function support on their worldvolumes, consistent with Abelian duality.  Their key assumption is that one promotes the currents to gauge field strengths in their own right, and they {\it assign}\ a kinetic term to them.  The resulting action is equivalent to (\ref{stuckjt}), and dual to the massive 2-form potential we start with. Our argument begins with an assumed form for the dual action, based on an analogy with the physical Higgs mechanism and specifically London's equation describing defect dynamics. The Julia-Toulouse theory (\ref{stuckjt}) follows from that.  At the level we and \cite{Julia:1979ur,Quevedo:1996uu,Quevedo:1996tx} work at, the dynamics describing the requisite proliferation of defects and their interactions that condense them is not given. For these details, one needs their underlying microscopic dynamics, in complete analogy to the BCS theory of the microscopic origins of superconductivity.  

\subsection{Comments on the Weak Gravity Conjecture}

While the nonlinearly realized gauge symmetry constrains corrections to the effective action, there are potentially additional constraints on the 4d field theory spectrum and the natural EFT cutoff, based on the so-called Weak Gravity Conjecture (WGC). The status of the WGC for massive gauge fields is, at present, unclear. We will confine ourselves to a discussion of some of the issues involved, and leave a direct assault on the question for the future.

The WGC, as formulated in \cite{ArkaniHamed:2006dz}, is a statement about $D$-dimensional effective field theories including gauge fields, with cutoff $\Lambda \lesssim m_{pl,D}$, coupled to gravity. It is an attempt to formulate a quantitative statement regarding quantum gravity-induced global symmetry breaking. It asserts that if one attempts to build a global symmetry by taking the gauge coupling $g_D$  to zero, the natural cutoff $\Lambda$ at which four-dimensional effective field theory breaks down will also vanish as a power of $g_D$. 

The sharpest arguments in \cite{ArkaniHamed:2006dz}\ stem from avoiding large numbers of stable electrically or magnetically charged black holes which could serve as remnants in the theory.  To avoid this, \cite{ArkaniHamed:2006dz}\ places bounds on the mass/tension of the lightest charged particles/p-branes, of the form
\be
	T^e_{p,D} \lesssim \sqrt{g_D^2 m_{pl,D}^{D-2}} \, ,
\ee
for $p$-branes electrically charged under $p+1$-forms in $D$ dimensions, and
\be
	T^m_{D-p-3,D} \lesssim \sqrt{\frac{m_{pl.D}^{D-2}}{g_D^2}} \, ,
\ee
for magnetically dual $D-p-3$-branes. In the latter case, the tension is typically a function of the cutoff of the effective field theory, so this places bounds on that cutoff. The point is that these bounds prevent one from taking the limit of vanishing charge, without simultaneously lowering the cutoff and thus the regime of validity of 4d effective field theory.

The basic argument in \cite{ArkaniHamed:2006dz}\ does not apply directly to massive gauge fields \cite{Cheung:2014vva}.  There are no black hole solutions which support such fields outside of the horizon, whether the mass is fundamental \cite{Bekenstein:1971hc,Bekenstein:1972ny,Teitelboim:1972qx}\ or emerges from spontaneous symmetry breaking \cite{Adler:1978dp}.\footnote{If the mass arises from a spontaneous symmetry breaking pattern that preserves a finite discrete symmetry, there is still potential quantum-mechanical hair \cite{Coleman:1991ku}. This type of hair is not associated with the classical long range fields which encode the conserved black hole charges, and it is not obvious that black holes with this kind of hair should lead to any limits on the low energy effective theory. We will sidestep this issue here.} Similarly, magnetic charge is confined in such theories, and a magnetically charged black hole must end on a magnetic flux tube which either stretches to infinity or ends on another magnetically charged black hole or a monopole of opposite sign.  One might therefore think that WGC might not have much to say about the spectrum of fields coupling via conserved currents to massive gauge fields.  

However, as the gauge field mass gets small enough, it is reasonable to ask how one would detect the mass in any physical process. For example, ref. \cite{Preskill:1990ty}\ claims that if a charged particle falls into a black hole, it will take a time scale of order $m^{-1}$, where $m$ is the gauge field mass, to discharge the massive electric field. For measurements on shorter time scales, the mass is essentially unobservable (by the energy-time uncertainty principle), and if the WGC constrains the massless theory on these time scales, it can be expected to yield at least some constraints on the massive theory as well, at least in the limit of small masses.  

While these questions are clearly interesting, we shall confine ourselves to a few largely heuristic comments about the massless limit of massive gauge fields which may be relevant for a deeper exploration of WGC-type constraints. 

For a massive gauge theory in Proca gauge, $m = e f$ with $f$ the periodicity of the Stueckelberg field. If the mass arrives from the spontaneous symmetry breaking in an Abelian Higgs model, $f$ is the vev of the radial mode of the charged scalar. There are two interesting massless limits.  One is to fix $e$ and send $f \to 0$.  In the context of the Abelian Higgs model, the symmetry becomes unbroken and a massless charged particle emerges. For finite $f$, if $e \, m_{pl} > m$, it seems plausible to require the existence of a light charged state with a mass $m$ which satisfies the WGC bound. If the Higgs self-coupling is $\lambda$, the Higgs field itself satisfies the constraint if $\lambda v^2 < e^2 m_{pl}^2$. Of course, once $v$ is small enough this will always be the case. All of this is consistent with the WGC conjecture: the Higgs mass is a threshold for the massive gauge theory, above which the gauge symmetry is unbroken and there is a light charged field; no additional charged matter is required.

A different perspective on this limit arises from considering the massive 2-form potential dual to the massive gauge field.  The dimensionful coupling of the two-form is $f$, and so this theory goes to weak coupling in the limit $f\to 0$, $m^2 \to 0$, consistent with the statement that the longitudinal mode of the 1-form gauge field is decoupling.  For $e \ll 1$, the mass $m = e f$ of $B$ is extremely light.  In the case that the dual theory is the low-energy limit of an Abelian Higgs model, the strings are the magnetic vortices whose tension scales as $f^2$, perhaps up to some logarithmic corrections. This is broadly consistent with a na\"ive application of the WGC would indeed require that electric strings which are charged under $B$ are becoming tensionless.

Another massless limit is $e \to 0$ with fixed $f$, which for the massless theory is conjectured to lead to the complete breakdown of 4d effective field theory coupled to gravity.  One possibility is that that renormalization effects could yield new nonperturbative vacua of the theory, which push the Higgs vev dynamically to extremely large values even for small charges, as discussed in \cite{Coleman:1973jx}. The gauge symmetry would remain broken, and the gauge field massive, screening the charges and possibly evading the violations of the  WGC bound.

Absent this possibility, we can again consider the limit from the viewpoint of the dual 2-form. The two-form gauge coupling $f$ is constant in our limit, while the mass $m = e f$.  In the strict limit, one has a massless two-form potential, which is dual to a massless scalar $\phi$ with field space periodicity $f$.  This theory has an apparent shift symmetry, which is supposed to be forbidden by quantum gravity. This can be avoided if instantons condense and break the symmetry. Gravitational instantons are believed to do so \cite{Giddings:1987cg,Abbott:1989jw,Kamionkowski:1992mf,Holman:1992us,Kallosh:1995hi}.  The dual statement is that the instanton condensate  generates a 4-form whose longitudinal mode is the 3-form field strength of the two-form potential. These instantons are magnetically charged under the 2-form potential; if that potential is massive, the instantons will be confined, bound together by Wilson lines. This transition, which is really a realization of the Julia-Toulouse proposal, and the role gravity plays in it, is an interesting subject for future work.

\section{Massive 3-form potentials}

We now turn to axion monodromy as realized by the dynamics of a massive 3-form potential, dual \cite{Dvali:2005an}\ to the axion-4-form description in \cite{Kaloper:2008fb,Kaloper:2011jz,Kaloper:2008qs,Aurilia:1980xj}.
As with the vector theory, the massive 3-form has good high-energy behavior, due in large part to the nonlinearly realized gauge symmetry, which stabilizes the theory against microscopic dynamics. In particular, the mass of the 3-form is radiatively stable, and higher powers of the potential $A$ are suppressed by higher powers of the gauge field mass $m$, much as with the massive vector.

The Julia-Toulouse mechanism \cite{Julia:1979ur,Quevedo:1996uu,Quevedo:1996tx} provides two routes to the required low-energy dynamics: via the condensation of membranes, inducing a mass for a fundamental 4-form and generating the inflaton as a condensate, or via a condensation of instantons which generates the 4-form as a collective mode that "eats" the two-form dual to the fundamental axion, opening up a mass gap and allowing the 4-form to drive inflation.  In these scenarios, the inflaton dynamics is that of a macroscopic condensate with energy densities well below the Planck scale: the size of the field in fundamental units simply counts the contributing sources in the condensate. 

A typical statement about large-field inflation is that it requires a softly broken shift symmetry, which gravity abhors.  The viewpoint which emerges from this paper is that there was never any continuous shift symmetry to protect. Rather, it is gauge invariance which determines the dynamics and protects the theory from Planck-scale physics. To drive this point home, we will revisit the dual description of the theory in terms of a periodic scalar coupled to a 4-form \cite{Kaloper:2008fb,Kaloper:2008qs,Kaloper:2011jz}. In this language, the theory is protected from UV corrections by a discrete gauge symmetry, which descends from the compactness of the gauge group of the initial massive 4-form.

The existence and onset of this massive phase must eventually be described within a well-defined microscopic theory, which sets the spectrum and dynamics of the defects, the mass of the gauge field, and so on. As field strengths increase, the microscopic theory will also determine whether some intermediate description is required, in which additional fields are activated, during inflation. Such phenomena might also lead to behavior such as the flattening of the inflaton potential that occurs in many string constructions \cite{Silverstein:2008sg,McAllister:2008hb,Dong:2010in}.  As with the one-form, there may be some general constraints on the spectrum of membranes coming from quantum gravity, as outlined in \cite{ArkaniHamed:2006dz}. At present we believe these arguments are not prohibitive, due in part to the physics of the massive phase.

\subsection{Lagrangian and dynamical scales}

Let us start with a massive 3-form U(1) gauge theory. In Stueckelberg formalism,
the basic Lagrangian with an $R_{\xi}$-style gauge-fixing term is:
\be
	{\cal L} = - \frac{1}{48} F_{\mu\nu\lambda\rho}F^{\mu\nu\lambda\rho} - \frac{m^2}{12} \left(A_{\mu\nu\lambda} - h_{\mu\nu\lambda}\right)^2 - \frac{1}{2\xi} \left(\p^{\mu} A_{\mu\nu\lambda} - \frac{\xi m^2}{2} b_{\nu\lambda}\right)^2 \, .\label{eq:fflag}
\ee
Here $F_{\mu\nu\lambda\rho} = 4 \p_{[\mu}A_{\nu\lambda\rho]}$ and $h_{\mu\nu\lambda} = 3 \p_{[\mu}b_{\nu\lambda]}$. We include the gauge-fixing term as we will be interested in the high-energy behavior of this theory, and following the discussion of the one-form, the properties of the propagator for $A$ will be crucial.  
The final gauge-fixing term is designed so that $\partial \cdot A$ and $b$ decouple, up to a boundary term. If we set $m b = B$, $m h = H$, so that $B$ has a canonical kinetic term, then\footnote{In principle we should not stop here. The `gauge transformation' two-form $B$ also has a gauge invariance $B \to B + d\Lambda^{(1)}$ which is broken only by the $\xi$-dependent mass term. So we should include a Stueckelberg one-form field $A$, transforming nonlinearly as $A \to A + \Lambda^{(1)}$, and a gauge-fixing term for this gauge invariance.  Finally, the field $A$ will have its own gauge invariance, and a gauge-fixing term and scalar Stueckelberg field should be added for this.  However, when $J$ is conserved, the original 3-form decouples from these Stueckelberg fields, and we will not worry about this complication.}
\be \label{4formmass}
	{\cal L} = - \frac{1}{48} F_{\mu\nu\lambda\rho}F^{\mu\nu\lambda\rho} - \frac{1}{12} \left(m A_{\mu\nu\lambda} - H_{\mu\nu\lambda}\right)^2 - \frac{1}{2\xi} \left(\p^{\mu} A_{\mu\nu\lambda} - \frac{\xi m}{2} B_{\nu\lambda}\right)^2 \, .
\ee
We can also couple $A$ to a 3-form current $J$.  The general gauge-invariant coupling is
\be
	\delta {\cal L}_J = (A - h)_{\mu\nu\lambda}J^{\mu\nu\lambda} = (A - h)\wedge {}^*J \, .
\ee
If $J$ is conserved, $d \,{}^*J = 0$, then the second term is a boundary term and $\delta {\cal L}_J = A\wedge {}^*J$.
Note, that if we consider the field equation for $F$ which follows from (\ref{4formmass}), and include the conserved 3-form current $\delta {\cal L}_J$ as a source we will find a nontrivial solution for the 3-form gauge field field potential $A$ even in the homogeneous static limit:
\be
A_{\mu\nu\lambda} \sim \frac{ J_{\mu\nu\lambda}}{m^2} \, ,
\ee
which is just the 3-form version of London's equation. Thus the physics is qualitatively similar to the case of a massive vector gauge field in a superconductor.

To understand the range of field space over which the theory (\ref{4formmass}) is defined, we can analyze the structure of the gauge transformations. 
Absent the gauge-fixing term, the above Lagrangian is invariant under the shift $A \to A + d\Lambda$, $b \to b +  \Lambda$.  If we assume that the volume of the gauge group is compact, this implies also that $b$ is a compact two-form gauge field.  More precisely, we assume as in \cite{Kaloper:2011jz}\ that the conjugate momentum for $A$ is quantized in units of the 3-form charge $q$. This also constrains the field $b$. 

To see this explicity, it is easiest to compactify the spatial directions on some $\Sigma^3$.  The dimensional reduction describes a particle on a circle. Starting with the assumption that the 4d momentum is quantized,
\be
	P_{123} = {\dot A}_{123} = n q \, ,
\ee
we can show that 
\be
	\int_{\Sigma^3} A \equiv \int_{\Sigma^3} A + \frac{2\pi}{q} \, .
\ee
This identification arises from a gauge transformation $A \to A + d\Lambda$ and uses the fact that the compactness of the gauge group implies  $\int_{\Sigma_2} \Lambda \to \int_{\Sigma_2} \Lambda + \frac{2\pi}{q}$ for any 2d submanifold $\Sigma_2 \subset \Sigma_4$. Effectively, $1/q$ is the radius of the group space circle. Since $b$ will shift by $\Lambda$, we similarly find 
\be
	\int_{\Sigma_2} b \equiv \int_{\Sigma_2} b + \frac{2\pi}{q} \, ,
\ee
which also implies that the magnetic $h$ flux is quantized:
\be
	\int_{\Sigma_3} h = \frac{2\pi n}{q}\label{eq:hfluxq} \, .
\ee

\subsection{High energy behavior}

A massless 3-form potential has no on-shell dynamics.  As is clear from the duality transformation above, the longitudinal mode of the massive 3-form {\it will} propagate. We will now argue that, so long as the 3-form couples to a conserved current, the theory will have good high-energy behavior just as a massive vector field does.

Following our discussion of the massive vector theory, we open a discussion of the high-energy behavior by studying the propagator of the massive 3-form.  A simple avenue to deriving this propagator is to write the 3-form potential as a dual vector:
\be
	A_{\mu\nu\lambda} = \epsilon_{\mu\nu\lambda\rho} V^{\rho}  \, .
\ee
In terms of $V^{\mu}$, the Lagrangian becomes
\be
	{\cal L} = - \frac{2}{\xi} \left(- \frac{1}{4} G_{\mu\nu}G^{\mu\nu} - \frac{\xi m^2}{4} V^2 - \frac{\xi}{4} (\p\cdot V)^2\right) \, ,
\ee
where $G$ is just the field strength for $V$. Amusingly, the 3-form kinetic term takes the form of a gauge-fixing term with respect to the dual one-form, and vice-versa.

The propagator for $V$ is thus a rescaled propagator for a vector field with mass $m \sqrt{\xi/2}$ and a gauge-fixing parameter ${\tilde \xi} = 2/\xi$.  The propagator for the 3-form potential is:
\be
	\langle A_{\mu\nu\lambda}(p) A_{\mu'\nu'\lambda'}(-p) \rangle  = 
		 \epsilon_{\mu\nu\lambda\rho}\epsilon_{\mu'\nu'\lambda'\rho'}\left( \frac{\frac{\xi}{2}\eta^{\rho\rho'}}{p^2 - \frac{\xi m^2}{2}} + \frac{\left(1 - \frac{\xi}{2}\right) p^{\rho}p^{\rho'}}{(p^2 - m^2)(p^2 - \frac{\xi m^2}{2})}\right)\, .
\ee
As with the vector field, there are no terms which grow large as $p \to \infty$ or $m \to 0$. Therefore as long as the 3-form couples only to conserved 3-form currents, the longitudinal and Stueckelberg forms will decouple. Thus as in the case of the massive vector theory, the amplitudes with the intermediate 3-form potential will have good high-energy behavior.

As with the massive vector theory, the gauge field mass will not receive any divergent counterterms, if the 3-form current is conserved and the matter sector which generates it has good high-energy behavior.   Antisymmetry and Poincar\'e invariance implies the self-energy diagram will have the form
\be 
	\Pi_{\mu\nu\lambda,\alpha\beta\gamma}(p) = \epsilon_{\mu\nu\lambda\rho}\epsilon_{\alpha\beta\gamma\delta}\left(A \eta^{\rho\delta} + B p^{\rho} p^{\delta}\right) \, . \label{eq:3fse}
\ee
Again, since the massive gauge field by itself is free, the self-energy diagram will require at least one insertion of the 3-form current operator. If we demand that this current be conserved, $p^{\mu}J_{\mu\nu\gamma}(p) = 0$, this means 
\be
	p^{\mu} \Pi_{\mu\nu\lambda,\alpha\beta\gamma} = 0\ .
\ee
Applying this to (\ref{eq:3fse}), we find that the term proportional to $B$ satisfies this constraint automatically. Thus, we must set $A = 0$.  The self-energy term is proportional to the momentum alone. The only divergent counterterm will be a wavefunction renormalization for the 3-form, and if the current sector has good high-energy behavior we expect the divergent counterterm to be at worst logarithmic.

\subsection{The dual axion and the Julia-Toulouse mechanism}

The massive 3-form theory is dual \cite{Dvali:2005an,Marchesano:2014mla}\ to the axion-4-form theory of \cite{Kaloper:2008fb,Kaloper:2008qs,Kaloper:2011jz}.   
We will review the duality map in detail, to make contact with \cite{Julia:1979ur,Quevedo:1996tx,Quevedo:1996uu}.

We start with the Lagrangian
\be \label{4formaction}
	{\cal L} = - \frac{1}{48} F_{\mu\nu\lambda\sigma}^2 - \frac{m^2}{12} (A _{\mu\nu\lambda} - h_{\mu\nu\lambda})^2 + \frac{m}{6} \phi \epsilon^{\mu\nu\lambda\rho}\p_{\mu} h_{\nu\lambda\rho} \, .
\ee
Here $\phi$ is a compact scalar, with periodicity $\phi \equiv \phi + 2\pi f$ to be determined.  
Integrating out $\phi$, we find $d h = 0$.  Locally, this means that we can set $h = db$. We can do this consistently in local coordinate patches if we identify $b$ under gauge transformations $b \to b + d\Lambda^{(1)}$.\footnote{If space is not simply connected, then as with the 2d path integral arguments for T-duality \cite{Buscher:1987qj,Rocek:1991ps}, we should take care to sum over winding sectors of $\phi$. This will enforce flux quantization for $h$.} The field $b$ becomes the Stueckelberg field in (\ref{4formmass}).  If we absorb $b$ into $A$ via a gauge transformation, we arrive at the Proca-like Lagrangian ${\cal L} = - \frac{1}{48} F^2 - \frac{m^2}{12} A^2$. Following our review of the massive vector, we will not consider more general $R_{\xi}$-like gauges when studying the duality transformations. We {\it will} use such gauges to better understand the high-energy behavior of the massive 4-form: as ever, different gauges are useful for making different aspects of the physics manifest.

We pass to a dual form of the action by integrating out $h$. We can complete the square in (\ref{jt3}), and perform the Gaussian integral, which is equivalent to solving the classical equations of motion for $h$, $h_{\mu\nu\lambda} = A_{\mu\nu\lambda} + \frac{1}{m} \epsilon_{\mu\nu\lambda\rho}\p^{\rho} \phi$, and substituting the solution back into (\ref{4formaction}). In this case we arrive at the axion-4-form theory described in \cite{Kaloper:2008fb,Kaloper:2011jz}:
\be \label{fform}
	{\cal L}_{\phi,A} = - \frac{1}{48} F_{\mu\nu\lambda\sigma}^2 - \half (\p_\mu\phi)^2 + \frac{m}{24} \phi \, \epsilon^{\mu\nu\lambda\rho}F_{\mu\nu\lambda\rho} \, .
\ee
The periodicity condition $\mu f = q$ derived in \cite{Kaloper:2011jz}\ arises naturally from this map.  The field redefinition $h_{\mu\nu\lambda} = A_{\mu\nu\lambda} + \frac{1}{m} \epsilon_{\mu\nu\lambda\rho}\p^{\rho} \phi$
directly implies $p_{\phi} = m h_{123}$, where $h_{123}$ is the spatial polarization for $h$.  The periodicity relation (\ref{eq:hfluxq}) then implies $m f = q$.

Proceeding in analogy to \S2.4, we wish to find a dual to the 4-form.  This is a somewhat odd step: by analogy, the dual would a zero-form field strength $q$, and there is no candidate potential that this would couple to. Furthermore, in these duality chains, the dual field appears as a Lagrange multiplier enforcing the Bianchi identity.  This identity is automatic for a 4-form.  Nonetheless, we introduce $Q$ as a Lagrange multiplier enforcing the equation $F = dA$ \cite{Kaloper:1993fg}, by treating $A^{(3)}$, $F^{(4)}$ as independent fields. 
We start with the Lagrangian 
\be \label{4formscalar2}
	{\cal L}_{\phi,A} = - \frac{1}{48} F_{\mu\nu\lambda\sigma}^2 - \half (\p_\mu\phi)^2 + \frac{m}{24} \phi \, \epsilon^{\mu\nu\lambda\rho}F_{\mu\nu\lambda\rho} + \frac{Q}{24} \, \epsilon_{\mu\nu\lambda\sigma}(F^{\mu\nu\lambda\sigma} - 4 \p^\mu A^{\nu\lambda\sigma}) \, ,
\ee
with the coefficient of the last term chosen to make the following formulae more compact. Integrating over $Q$ enforces 
$F = dA$.  We allow $A$ in different coordinate patches to be related by nontrivial gauge transformations.

On the other hand we can integrate out $F$, by completing the square in the action, again equivalent to finding the equation of motion for $F$ as an independent field and inserting the result back into the action. This yields the equation $F_{\mu\nu\lambda\sigma} = (Q +m \phi) \epsilon_{\mu\nu\lambda\sigma}$, or if we invert it, 
$Q +m\phi = - \epsilon_{\mu\nu\lambda\sigma} F^{\mu\nu\lambda\sigma}/24$. Substituting these equations back in the action (\ref{4formscalar2}) yields
\be \label{4formscalar3}
	{\cal L}_{\phi,A} = - \half (\p_\mu\phi)^2 - \frac{m^2}{2} (\phi+ \frac{Q}{m} )^2  + \frac{1}{6} \, \epsilon_{\mu\nu\lambda\sigma} Q \, \p^{\mu} A^{\nu\lambda\sigma}\, .
\ee
Note that varying $A$ yields the equation of motion $d Q = 0$, consistent with the quantization condition. Furthermore, the compactness of the gauge group means that summing over integer units of flux $\int dA$ enforces a quantization condition $Q = 2\pi n q$. The resulting action is consistent with the discrete gauge invariance $\phi \to \phi + 2\pi f$ if $Q$ transforms as $Q \to Q - 2\pi q$ and $m f = q$.

The result is precisely the scalar field action of \cite{Kaloper:2008qs,Kaloper:2011jz}. Our chain of reasoning however allows a discussion of these actions in parallel with \S2.4, which we give here at a very heuristic level.  We can start with a compact scalar, defined as a scalar that shifts by a discrete gauge symmetry.  This can be given a mass if we add a ``Stueckelberg field" $Q$ which also transforms nonlinearly under this symmetry. Pursuing our analogy, the scalar $\phi$ couples as a gauge field to a charged instanton; $Q$ denotes the instanton condensate. Reversing the duality steps above, we find that the condensate dynamics can be described as those of a massive 3-form, with the scalar $\phi$ the dual of the 2-form longitudinal mode.  In this case, the massive 3-form gauge theory describes the dynamics for some kind of instanton condensate.  

Another route to this set of dual effective actions is to consider a 3-form potential which couples electrically to charged membranes.  It is natural to suppose that  if and when these membranes condense, the result is to give a mass to the 3-form, such that the two-form Stueckelberg field parametrizes the membrane condensate.  Pursuing the chain of dualities to the action (\ref{4formscalar3}), we claim, following Julia and Toulouse, that the inflaton dynamics is that of a membrane condensate. The analogue of the Julia-Toulouse ans\"atz (\ref{jtansatz})
for massive vector fields is the equation
\be
\phi' = - \frac{1}{24 m} \epsilon_{\mu\nu\lambda\sigma} F^{\mu\nu\lambda\sigma} \, ,
\ee
after the ``gauge" variable $Q$ has been absorbed into $\phi' = \phi + Q/m$. 
Note that -- by the discussion of the quantization of $Q$ above -- as this quantity is a dual of the $4$-form flux, it is a sum total of many different units of flux inside the volume occupied by the field configuration. So $Q$ is a macroscopic, extensive property of the system, whose large value reflects system's size and number of constituent parts rather than being a characteristic of a high energy excitation of the system.
Macroscopically large fluxes of $F_{\mu\nu\lambda\sigma}$ thus translate into huge initial displacements of $\phi'$ from its minimum without ever taking the theory out of the validity of the low energy description.

It is clearly of great interest to find UV complete models in which membrane condensation and the emergence of a dual scalar occurs as an infrared phenomenon.  This does occur in the two dimensional dimensional reduction of the axion-4-form theory, which is just a compact boson coupled to a 2d Abelian gauge field:
\be
	L = - \half F_{01}^2 - \half (\p\phi)^2 - \frac{\phi}{2\pi} F_{01} - \mu^2 \cos\phi\ .
\ee
A classic treatment can be found in \cite{Coleman:1976uz}.  This and related 2d models were studied in light of axion monodromy in \cite{Lawrence:2012ua}.  Charged fermions behave as domain walls; Bose-Fermi duality maps the scalar $\phi$ to  charged fermion with mass $\mu$, the latter being the 2d analog of a charged domain wall  \cite{Coleman:1976uz,Coleman:1975pw,Coleman:1974bu,Mandelstam:1975hb}. $\phi$ is a composite object from the fermionic point of view. In two dimensions this is an exact duality. In higher dimensions we would search for an infrared duality.

\subsection{Effective field theory analysis}

Next, we discuss the low energy effective field theory dynamics of the massive 3-form gauge theory. 
When we embed our 3-form theory in a UV-complete theory, we expect the duality between scalar and longitudinal mode to hold only in the infrared. In considering the effect of the UV completion, we should pick a duality frame in which to write down local couplings between the inflaton sector and additional fields. Thus, we will
re-run the discussion in \cite{Kaloper:2008fb,Kaloper:2011jz}\ in the duality frame of the massive 4-form. Once this is done, we can discuss the IR dynamics in whichever frame is convenient.

\subsubsection{Terms of the form $F^{2k}$}

One set of leading terms are, as in \cite{Kaloper:2008fb,Kaloper:2011jz}, higher powers of the 4-form field strength,
\be
	\delta {\cal L}_1 = c_n \frac{F^{2n}}{M^{4n-4}} \, 
\ee
(assuming parity is conserved in this sector). As in \cite{Kaloper:2008fb,Kaloper:2011jz}, these will lead to corrections to the inflaton potential of the form
\be
	\delta V \sim V \frac{V^n}{M^{4n}} \, .
\ee

In this discussion we treat the corrections as small perturbations of the leading order action, and consider the effective action of the theory as a series in higher dimension operators. There are known cases in which higher order terms resum to a function that can be analytically continued outside of the radius of convergence of the effective field theory expansion: a classic example is \cite{Coleman:1973jx}. This may come from integrating out additional fields, which can lead for example to a ``flattened" potential away from the minimum at $F = A = 0$ \cite{Silverstein:2008sg,McAllister:2008hb,Dong:2010in}. For such cases, we recall the following treatment in \cite{Dvali:2005an}.
Start with the general action of the form:
\be \label{4formactiongen}
	{\cal L} =  M^4  K\Bigl(\frac{F}{M^2}\Bigr) - \frac{m^2}{12} (A _{\mu\nu\lambda} - h_{\mu\nu\lambda})^2
	\, ,
\ee
where we use the fact that a 4-form in four dimensions has only one independent component, $F = - F_{\mu\nu\lambda\sigma} \epsilon^{\mu\nu\lambda\sigma}/24$, and $F_{\mu\nu\lambda\sigma}^2 = - 24 F^2$. 
We can still dualize this action as before, to find the Lagrangian for the scalar $\phi$.  
The resulting axion effective potential is, using $z= (Q+m\phi)/M^2$, 
\be \label{potmonaxgen}
V\Bigl(z\Bigr) = M^4 \Bigl\{z W(z) - K\Bigl(W(z) \Bigr) \Bigr\} \, ,
\ee
where $W = (K')^{-1}$.  Thus the scalar axion potential is just the Legendre transform of the kinetic term of the 3-form, and the magnetic dual $Q$ of the 4-form field strength provides the monodromy structure. Under this map, a quadratic function in $F$ yields a quadratic function in $z$. 

\subsubsection{Terms of the form $A^{2k}$}

The next set of terms which can appear takes the form
\be
	\delta {\cal L}_2 = \frac{d_n}{M^{2n-4}} (A - h)^{2n}\label{eq:highergv} \, .
\ee
The duality map above takes this to $\delta {\cal L}_2 = \frac{d_n}{m^{2n} M^{2n - 4}} (\p\phi)^{2n}$.
If such terms were induced by fields with mass $M$ coupling directly to $\phi$, we would expect that $d_n \sim \left(\frac{m}{M}\right)^{2n}$, so that the $m^2 \to 0$ limit is smooth. We wish to argue that this scaling also holds if the dual massive 4-form is coupled to matter with the good high-energy behavior a renormalizable QFT or string theory would provide. 

To this end, gauge invariance points us to a 3-form version of the Goldstone Boson Equivalence Theorem: the scattering of the longitudinal mode of $k$ 3-form gauge bosons should be of order $m^k/E^k$ times the scattering of $k$ canonically normalized 2-form gauge fields, up to terms of still higher order in $m^2/E^2$. This is obvious in a way since the massive 4-form only has the longitudinal modes propagating, being locally a constant in the massless limit. Gauge invariance requires powers of $A$ without derivatives in the effective action to come in the form 
$m (A - h) = mA - H$. Since the massive 3-form has good high-energy behavior, $m$ does not function as a strong coupling scale and we expect the scattering of canonically normalized two-form to be nonsingular in the $m^2 \to 0$ limit, which implies our claim that $d_n = d'_n m^{2n}/M^{2n}$.  Thus the dual of (\ref{eq:highergv}) are higher-derivative axion couplings of the form
\be
	{\widetilde {\delta {\cal L}}}_2 \sim \frac{d'_n}{M^{4n-4}} \left(\p\phi\right)^{2n} \, .
\ee
This gives a negligible contribution to density fluctuations generated by inflation if $M \gg H$, where $H = \sqrt{V}/\sqrt{} m_{pl}$ is the Hubble scale during inflation \cite{Kaloper:2002uj,Kaloper:2011jz}.

Further possibilities arise form terms of the form $A^{2k} F^{2m}$. We expect the above arguments to yield coefficients that scale as $m^{2k}$.  The translation of these terms to inflaton dynamics is a matter for future work.

\subsubsection{Inflation models and flattening}

If the UV scale is above $M_{GUT}\sim 2\times 10^{16}\ GeV$, and the 3-form sector does not couple to moduli with masses below the inflationary Hubble scale and Planck-suppressed couplings to the inflaton, then as in \cite{Kaloper:2008fb,Kaloper:2011jz}, we arrive at a model of inflation nearly identical to  the classic $\half m^2 \phi^2$ chaotic inflation.  As we argued in \cite{Kaloper:2014zba}, finding such a model may require stringy compactifications for which the 10 or 11d SUGRA approximation does not apply. Even in these models, there is ample room for interesting observable UV-sensitive corrections to the vanilla $\half m^2 \phi^2$ model. Particularly interesting would be the corrections suppressed by scales close to the scale of inflation, since in this case the corrected potential can lead to observably different signatures than the plain vanilla $\half m^2 \phi^2$. These theories benefit from living dangerously close to the limits of the validity of the perturbative description. 

On the other hand, in the original constructions of axion monodromy via string theory \cite{Silverstein:2008sg,McAllister:2008hb,Dong:2010in,McAllister:2014mpa}, and in the field theory construction \cite{Dubovsky:2011tu}, the potentials flatten considerably away from the quadratic form. For example potentials written in terms of the canonically normalized inflaton can scale as $V \sim \phi^{k < 2}$. In our language this indicates the dominance of operators such as $F^{2k}$ with very high powers of $k$, indicating field strengths $F$ above the dynamical scale $M$ generating this potential.  Whether such terms are calculable, following a treatment such as\cite{Coleman:1973jx}, depends on the dynamics at the scale $M$.

It is worth revisiting the model in the introduction of \cite{Dong:2010in}, in our framework.  As \cite{Dong:2010in} we will take it as a model of how a flattened potential might arise, without attempting to complete it into a full theory of inflation. Consider an ``inflaton" $\phi$ and a heavy field $\psi$, with potential
\be 
	V = \half g \psi^2 \phi^2 + \half \mu^2 (\psi - \psi_0)^2 \, .
\ee
Following \cite{Dong:2010in}\ we demand that $\psi_0 \ll m_{pl}$, so that Planck-suppressed operators will not unduly affect the dynamics of $\psi$. We can rewrite this in our language:
\be
	L \sim \frac{1}{48} \left(1 - \frac{\alpha \psi}{m_{pl}}\right) F^2 - \frac{m^2}{12} A^2 - \half \mu^2 \psi^2 \, ,
\ee
where we set $\psi_0 \equiv m_{pl}/\alpha$, $\alpha \gg1$.  If we integrate out $\psi$ at tree level, we find
\be \label{eq:hpot}
	L(F) = f\left(\frac{\alpha^2 F^2}{m_{pl}^2 \mu^2}\right) F^2
\ee
where $f(x) \sim 1 + {\cal O}(1) x$ for $x \ll 1$, and $f(x) \sim 1/x$ for $x \gg 1$.  The relevant dynamical scale controlling which regime the theory is in is
\be
	M^4 \equiv m_{pl}^2 \mu^2/\alpha^2 = \mu^2 \psi_0^2 \, .
\ee
The scale $M$ sets the value of $F$ at which the potential begins to flatten, as well as the energy density in this regime.

In \cite{Kaloper:2011jz,Kaloper:2014zba}\ we noted that a quadratic monodromy potential required that any moduli coupling to the inflaton have masses $\mu \gg H$, where $H$ is the Hubble scale during inflation.
As noted in \cite{Dong:2010in}, to avoid flattening in (\ref{eq:hpot}), we must take $\mu \gg \alpha H$, where $H^2 = V/3 m_{pl}^2$. Note that super-Hubble masses can still lead to flattening. The point is that in \cite{Kaloper:2011jz,Kaloper:2014zba}\ we defined moduli as scalar fields with Planck-suppressed couplings, in which case $\alpha \sim 1$.  Which scenario is more likely in string theory is a question we do not address here. 

Such potentials, with new scalar modes activated, may arise if we formulate the analog of the Ginzburg-Landau equation for our superconducting phase.  It would be interesting and useful to construct  such``intermediate"-level effective theories yielding monodromy inflation with flatter-than-quadratic potentials, and explore their UV sensitivity.

\subsection{Discrete gauge invariance and axion monodromy}

In this subsection we turn more directly to the quantum mechanics of the dual description of the massive 3-form \cite{Kaloper:2008fb,Kaloper:2008qs,Kaloper:2011jz}, to shed some light on the physics that protects large field displacements from quantum gravity effects. As we noted above, the axion-4-form theory has a discrete gauge symmetry $\phi \to \phi  + 2\pi f$. In the Lagrangian description, this protects it from large UV corrections. In the Hamiltonian language the gauge symmetry is hidden \cite{Kaloper:2011jz}. The canonical momentum for the 4-form flux is quantized in units of the 4-form charge $q$ as $p = *F^{(4)} - m\phi = n q$, with $n \in \ZZ$. The Hamiltonian is
\be
	H = \half p_{\phi}^2 + \half \left(m \phi + n q\right)^2 \, .
\ee
It is periodic under $\phi \to \phi + 2\pi f$ if $2\pi m f = k q$, with $k \in \ZZ$, and if we shift $n \to n - k$. Thus we see that the discrete gauge symmetry acts on phase space of the theory.

To clarify this, let us consider a massive charged particle in two dimensions, coupled to a magnetic field,
\be
	L = \half {\dot x}^2 + \half{\dot y}^2 - g y {\dot x}\, ,\label{eq:qmmag}
\ee
where we have set the mass $m = 1$. Eq. (\ref{eq:qmmag})\ is the dimensional reduction of the axion-4-form model of \cite{Kaloper:2008fb,Kaloper:2011jz}\ to zero spatial dimensions, encapsulating the nontrivial dynamics of the system. The `coordinate' $y$ is the axion zero mode, while $x$ is the 3-form gauge potential.  
We further place the particle on the torus $T^2$, so that $x,y$ are periodic with periods $2\pi R_{x,y}$. 
From the viewpoint of our original 4d axion-4-form theory, the periodicity arises when the original axion is a compact scalar, and the gauge group of the 3-form theory is a compact $U(1)$, so that Wilson hypersurfaces $\int_{\Sigma_3} A$ are compact.

The periodicity of $T^2$ is a genuine discrete gauge symmetry of the theory. The term proportional to $g$ is invariant up to a total derivative. The subtlety is to ensure that this gauge symmetry is consistent with quantization. Doing so imposes a quantization of $g$ in units of $1/(2\pi R_x R_y)$. Indeed, the canonical momenta are:
\be	p_x = {\dot x} - g y = \frac{k}{R_x} \, , ~~~~~~~~~~~~~ 
	p_y = {\dot y} \, .
\ee
Since $x \to x + 2\pi R_x$ is a manifest discrete gauge symmetry, the wavefunctions must be invariant under it, and so the conjugate momentum $p_x$ will be quantized. On the other hand, if $y \to y + 2\pi R_y$, then $p_x \to p_x - 2\pi g R_y$, which is only consistent with the quantization condition on $p_x$ if $g = \frac{\ell}{2\pi R_x R_y}$ for $\ell \in \ZZ$. This is the analog of the condition on mass, axion decay constant, and 4-form charge given in Eq. (6) of \cite{Kaloper:2011jz}. Furthermore, it is clear that the gauge symmetry corresponding to shifts of $y$ acts nontrivially on phase space.  

The Hamiltonian can then be rewritten as
\be
	H = \half (p_x + g y)^2+ \half p_y^2 \, ,
\ee
where the index $\ell = 2\pi R_x R_y g$ is the number of magnetic flux quanta, seting the degeneracy of a Landau level.  For $\ell = 1$, that degeneracy is unity, and the energy of the $n$th Landau level is $g (n + \half) $ up to some factors.  This is the choice made in \cite{Kaloper:2008fb,Kaloper:2011jz}.  The Hamiltonian is just the linear (aka simple) harmonic oscillator or LHO. One can create a coherent state out of LHO creation and annihilation operators, which is to all intents and purposes describing the oscillator high up on the side of the potential. Such a large excitation, counted by the high occupation number of the LHO models precisely the large inflaton {\it vev}. 

For a higher number $\ell$ of flux quanta, the degeneracy of the Landau levels increases.  In the axion-4-form story, this arises from the M-theory compactification of \cite{Kaloper:2008fb}. That is, consider a compactification on $M_7 = T_4\times T_3$ or on some $K_3$ fibration of a 3d base $\Sigma$, perhaps on a manifold $G_2$ holonomy. The 3-form potential has a Wilson surface through the $T_3$ or $\Sigma$, leading to a 4d axion.  If one sets $\int_{T^4\ {\rm or}\ K3} F^{(4)} = \ell$, one arrives at the 4d analog of the case of degenerate Landau levels.

The phase space in the $p_x-y$ direction is a tilted cylinder, and so $y$ is no longer periodic by itself in the phase space. For a fixed value of $k$, $k/(g R_x)$, the lowest Landau level looks like the minimum of the LHO, and this minimum should sit inside the torus ({\it cf.} \cite{Tong:2016kpv}).  However the assertion that the Landau levels are matched to the states of an LHO presumes that one ``unpacks" the coordinate $y$ by removing the excess units of the period to fit the value in the fundamental domain: $y$ larger than $2\pi R_y$ is equivalent to $y$ inside the fundamental domain, but with $k$ shifted by $\ell$. For $\ell = 1$, this gives the picture in Figure 1 of \cite{Kaloper:2011jz}.

The structure of quantized phase space restricts the corrections to the zero mode dynamics, which in turn protects the inflationary slow roll. In the quantum-mechanical theory, terms $\sim y^n$ cannot be added to the Hamiltonian as they are not periodic in phase space. They must be completed into the form $c_n (p_x + g y)^n = c_n {\dot x}^n$.  Lifting to 4 dimensions, $y \to \phi$, ${\dot x} \to F^{(4)}$, this means that the terms in the Hamiltonian that involve $\phi^k$ must be completed into $(p_{A^{(3)}} + m \phi)^k$ where $m$ is the mass term. But this combination is just $(F^{(4)})^k$.

The guideline of the effective field theory analysis of \cite{Kaloper:2008fb,Kaloper:2011jz}\ was that the UV completion is such that corrections to the low-energy effective action take the form ${\cal O}(1) (F/M_{UV}^2)^{2k}$, without powers of the gauge field mass $m$ in the denominators. As we have argued above, this follows from fairly conservative assumptions about the low energy theory and its UV completion, based on locality, unitarity and renormalizability. Under these conditions large-field inflation can be made safe from dangerous corrections so long as $M_{UV}$ is a little bit larger than the scale of inflation \cite{Kaloper:2008fb,Kaloper:2011jz}.

\subsection{The Weak Gravity Conjecture and membranes}

The 4-forms couple electrically to charged membranes, whose nucleation induces jumps in flux of order $n q$. So the viability of inflation requires that these membranes are not so light that their emission `discharges' the initial field displacement too quickly, spoiling slow roll
\cite{Kaloper:2008fb,Kaloper:2011jz,Kaloper:2014zba}. It appears reasonable to assume that the WGC places an upper bound on such charged membranes, though opinions differ as to how prohibitive this constraint is \cite{Brown:2015iha,Ibanez:2015fcv,Hebecker:2015zss,Brown:2016nqt}. 

However, it is not entirely clear that the arguments in \cite{ArkaniHamed:2006dz}\ should directly apply to codimension one charged objects, as discussed in \cite{Ibanez:2015fcv}. 
Furthermore, the 3-form gauge theory we describe is massive, preventing a direct application of the arguments in \cite{ArkaniHamed:2006dz}.  

In the massive 3-form case, we have argued that the inflaton can be considered as parametrizing a membrane condensate.  A related argument appears in section 2 of \cite{Hebecker:2015zss}.  One can consider instanton-induced sinusoidal modulations of the effective potential for $\phi$.  The associated domain walls between minima of this sinusoid can be light, depending on the strong coupling scale of the gauge theory producing the potential. If we can dial this scale, eventually the scalar is no longer trapped and can roll classically over the sinusoid.  At this point the inflaton is behaving as a condensate of membranes, in line with the Julia-Toulouse mechanism \cite{Julia:1979ur,Quevedo:1996uu}. 

It is tempting to speculate, as do \cite{Hebecker:2015zss}, that the inflaton thus automatically satisfies the ``electric" form of the WGC.  We should be a little careful.  Recall that for massive vectors, initial gauge fields emanating from a black hole are completely screened by the condensate over a time scale of order $m^{-1}$, with $m$ the gauge field mass.  When this mass becomes small, the status of this ``electric" form is unclear, as it can depend on the UV completion of the massive gauge theory.  For the massive 3-form, the scalar discharges the 4-form flux, with a time scale set by its inverse mass, which is also the inverse mass of the gauge field.   

On the other hand, an argument following \cite{Hebecker:2015zss} may lead to an upper bound on the validity of effective field theory, albeit one weak enough to allow for inflation. So long as the WGC holds in higher dimensions (for which the domain wall may have higher codimension), the 4d theory may not have domain walls with tension $T = M_{WGC}^3$ satisfying the na\"ive lower bound given in \cite{ArkaniHamed:2006dz}, if the 4d effective field theory breaks down at a scale below $M_{WGC}$.  This can arise via a low Kaluza-Klein scale, $\Lambda_{KK} < (q m_{pl})^{1/3}$, where $\Lambda$ is the scale at which 4d effective field theory breaks down.  As pointed out in \cite{Hebecker:2015zss}, inflation can proceed so long as this scale is above the Hubble scale. 
One may also worry that terms of the form $F^{2k}$ would be suppressed only by the KK scale, imposing a stronger constraint. However, Kaluza-Klein modes come with $m_{pl}$-suppressed couplings ({\it cf.} \cite{Kaloper:2002uj,Kaloper:2014zba}). Properly summing over Kaluza Klein modes leads to such operators suppressed by the higher-dimensional Planck scale, so that the constraint is still $\Lambda_{KK} > H$. All that is required is thus that the higher-dimensional Planck scale be large compared to the energy scale at which inflation operates. 

\section{Conclusion}

CMBR observations will soon be able to efficiently constrain inflationary models at highest scales, or, with luck, discover them. This forces the question of the internal self-consistency of inflationary models.
A UV complete model provides a proof in principle of such consistency, as well as a model with explicit, computable parameters. However it is clearly of interest if one can make general arguments as to which -- if any -- large field models are self-consistent, and why.

Here we have pursued this question from the effective field theory point of view, and argued that massive abelian 4-forms provide robust effective field theories realizing axion monodromy inflation. Instead of a broken global shift symmetry, which is expected to be badly broken by quantum gravity, there is a nonlinearly realized gauge symmetry for the 3-form, and an associated discrete gauge symmetry for the dual inflaton. These control the possible UV corrections. We know of no principle in quantum gravity which prevents these gauge symmetries.  

Further, the large field excursions by the effective inflaton are duals of large gauge field fluxes. These large fluxes -- and in turn, the effective inflaton field values -- can be thought of as macroscopic quantities which characterize the size of the system, rather than high energy excitations of the inflaton. While there is an upper limit on just how large a flux can be, this would imply that it is controlled by the macroscopic properties of the low energy theory, rather than its UV features.

The major question is whether the 3-form mass can be kept well below the Planck scale.  In string theory models with high-scale supersymmetry, there will be $p$-form gauge symmetries that become linearly realized in the ultraviolet, which can become nonlinearly realized at a lower scale.  The gauge mass and the presence of extra modes that may activate flattening of the inflaton potential are set by the string compactification and the IR dynamics that emerge. If the Julia-Toulouse mechanism \cite{Julia:1979ur,Quevedo:1996uu,Quevedo:1996tx}\ can be realized as IR dynamics in a UV-complete theory, this may provide an avenue to generate the needed scale hierarchies. Full string compactifications are one route to this.  It would be interesting to find intermediate-scale four-dimensional models of Ginzburg-Landau type, in which one can see the transition from a massless to a massive gauge theory phase.  Higher-dimensional intermediate models, such as the ``unwinding inflation" models of \cite{DAmico:2012ji,DAmico:2012sz}\ are also of great interest.\footnote{The $D3-{\bar D3}$ version in five dimensions has a clear relation to the present work.  Here the D3-brane separation on the circle is the inflaton.  If we T-dualize, the off-diagonal Wilson line dual to the D4 has a Chern-Simons coupling to the RR 4-form field strength, which upon compactification becomes the axion-4-form coupling we describe here.}


\vskip.5cm

{\bf Acknowledgments}: 
We would like to thank Guido D'Amico, Daniel Harlow, David E. Kaplan, Matthew Kleban, Emil Martinec, Hirosi Ooguri, Fernando Quevedo, Eva Silverstein and Lorenzo Sorbo for useful conversations on this and related subjects. N.K. thanks the CERN Theory Division and Particle Theory Group, University of Nottingham, for hospitality in the course of this work. Part of this work was carried out while A.L. attended the Aspen Center for Physics during the ``Primordial Physics" workshop, which is supported by National Science Foundation grant PHY-1066293.  A.L. would like to thank the organizers, participants, and the ACP staff for a stimulating environment conducive to work. Part of this work was carried out while A.L. attended the ``Entanglement in Strongly-Correlated Quantum Matter" and ``Quantum Gravity Foundations: UV to IR" workshops at the KITP, during which his research was supported in part by the National Science Foundation under Grant No. NSF PHY11-25915. He would like to thank the organizers, participants, and KITP staff for a stimulating environment. N.K. is supported in part by the DOE Grant DE-SC0009999. A.L. is supported in part by DOE grant DE-SC0009987.

\bibliographystyle{utphys}
\bibliography{london_refs.bib}

\end{document}